\documentclass[prd,aps,floatfix,nofootinbib,11 pt]{revtex4}
\usepackage{amsmath,graphicx,color,epsfig}

\input{tcilatex}

\begin{document}

\title{Dynamics of multipartite quantum correlations under decoherence}
\author{M. Ramzan\thanks{%
mramzan@phys.qau.edu.pk}}

\address{Department of Physics Quaid-i-Azam University \\
Islamabad 45320, Pakistan}

\date{\today}

\begin{abstract}
Quantum discord is an optimal resource for the quantification of classical
and non-classical correlations as compared to other related measures.
Geometric measure of quantum discord is another measure of quantum
correlations. Recently, the geometric quantum discord for multipartite
states has been introduced by Jianwei Xu [arxiv:quant/ph.1205.0330].
Motivated from the recent study [Ann. Phys. 327 (2012) 851] for the
bipartite systems, I have investigated global quantum discord (QD) and
geometric quantum discord (GQD) under the influence of external environments
for different multipartite states. Werner-$GHZ$ type three-qubit and
six-qubit states are considered in inertial and non-inertial settings. The
dynamics of QD and GQD is investigated under amplitude damping, phase
damping, depolarizing and flipping channels. It is seen that the quantum
discord vanishes for $p>0.75$ in case of three-qubit $GHZ$ states and for $%
p>0.5$ for six qubit $GHZ$ states. This implies that multipartite states are
more fragile to decoherence for higher values of $N$. Surprisingly, a rapid
sudden death of discord occurs in case of phase flip channel. However, for
bit flip channel, no sudden death happens for the six-qubit states. On the
other hand, depolarizing channel heavily influences the QD and GQD as
compared to the amplitude damping channel. It means that the depolarizing
channel has the most destructive influence on the discords for multipartite
states. From the perspective of accelerated observers, it is seen that
effect of environment on QD and GQD is much stronger than that of the
acceleration of non-inertial frames. The degradation of QD and GQD happens
due to Unruh effect. Furthermore, QD exhibits more robustness than GQD when
the multipartite systems are exposed to environment.\newline
\end{abstract}

\maketitle

\newline

\vspace*{1.0cm}

\vspace*{1.0cm}



\section{Introduction}

During recent years, quantum discord [1-14] has become the main focus of
fundamental research in the discipline of quantum information theory. It
quantifies the total non-classical correlations in a quantum state.
Recently, quantum discord for multipartite system has been investigated
[15-17]. Its extension to $N$-partite $GHZ$ state is given by Jianwei Xu
[18]. The dynamics of quantum discord has been extensively studied in
different contexts by various authors [19-45]. It has been shown that
quantum discord is more robust than entanglement for bipartite systems.
Furthermore, the quantum discord for a qubit-qutrit [46] and qubit-qudit
[47] systems has been proposed. Quantum discord has been used in studies of
quantum phase transition [48, 49] and to measure the quantum correlation
between relatively accelerated observers [50]. The geometric interpretation
of the geometric discord has been discussed by Yao et al. [51]. The lower
and upper bounds of quantum discord have also been investigated [52, 53].
Its experimental evidences can be seen in references [54-57].

Recently, the geometric measure of quantum discord (GMQD) has been proposed
[58, 59], which quantifies the amount of non-classical correlations of a
state in terms of its minimal distance from the set of genuinely classical
states. However, most of the studies in this connection deal with bipartite
quantum states. It has also been studied over two-sided projective
measurements by [60]. Recently, the geometric quantum discord (GQD) for
multipartite quantum states have been introduced by Jianwei Xu [61] and its
lower bound is given. Motivated from the previous studies regarding the
influence of environment on the bipartite quantum discord [62] and its
geometric measure [63], I have investigated the global quantum discord (QD)
and its geometric measure (GQD) under the influence of external environments
for different types of multipartite quantum states. Quantum systems can
never be isolated from their environments completely and the interactions
with the environment deteriorate the purity of the quantum states. This
gives rise to the phenomenon of decoherence [64], which appears when a
system interacts with its environment in a irreversible way. It plays a
fundamental role in the description of the quantum-to-classical transition
[65] and has been successfully applied in the cavity QED [66]. Another
familiar aspect, the degradation of entanglement has also been investigated
recently by several authors [67-70], with special attention from the
non-inertial perspective. Entanglement in noninertial frames was first time
introduced by Alsing et al. [71]. The subject have attracted much attention
during recent years [72-80]. It has also been investigated under decoherence
for bipartite [81-85], qubit-qutrit [86] and multipartite [87] systems. The
entanglement dynamics for noninertial observers in a correlated environment
is considered in Ref. [88], where it is shown that correlated noise
compensates the loss of entanglement caused by the Unruh effect. Recently, I
have studied the decoherece dynamics of GMQD and MIN at finite temperature
for non-inertial observers [89].

In this paper, I have investigated the decoherence dynamics of multipartite
quantum discord and geometric quantum discord for $GHZ$-type initial states
in inertial and non-inertial frames. Different decoherence channels are
considered parameterized by decoherence parameter $p$ such that $p\in
\lbrack 0,1]$. The lower and upper limits of decoherence parameter represent
the fully coherent and fully decohered system, respectively. It is seen that
the quantum discord is more robust than geometric quantum discord under
environmental effects. However, Werner-$GHZ$ type states are found to be
more fragile to decoherence for higher values of $N$. The depolarizing
channel heavily influences the QD and GQD as compared to the amplitude
damping channel. It is also seen that the effect of environment on QD and
GQD is much stronger than that of the acceleration of the accelerated
observer. The degradation of QD and GQD happens due to Unruh effect.

\section{Evolution of multipartite quantum states}

The quantum discord, a measure of the minimal loss of correlation in the
sense of quantum mutual information, can be defined for a bipartite quantum
state $\rho _{AB}$ composed of subsystems $A$ and $B$, as [1]%
\begin{equation}
QD(\rho _{AB})=I(\rho _{AB})-C(\rho _{AB})
\end{equation}%
where%
\begin{equation}
I(\rho _{AB})=S(\rho _{A})+S(\rho _{B})-S(\rho _{AB})
\end{equation}%
is the quantum mutual information and measures. Here $\rho _{A,B}=$Tr$%
_{B,A}\rho _{AB}$ are the reduced density matrices and
\begin{equation}
S(\rho )=-\text{Tr}(\rho \log \rho )
\end{equation}%
is the von-Neumann entropy. $C(\rho _{AB})=\max_{\{\Pi \}}[S(\rho
_{A})-S(\rho _{AB}|\{\Pi _{k}\})]$ is the measure of classical correlations
between the two subsystems. It is defined as the maximum information about
one system that can be obtained by performing a set of projective
measurements on the other subsystem and the maximum is taken over the set of
projective measurements $\{\Pi _{k}\}$ as%
\begin{equation}
S(\rho |\Pi ^{A}):=\sum\limits_{k}p_{k}S(\rho _{k})
\end{equation}%
and
\begin{equation}
\rho _{k}=\frac{1}{p_{k}}(\Pi _{k}^{A}\otimes I_{B})\rho (\Pi
_{k}^{A}\otimes I_{B})
\end{equation}%
with
\begin{equation}
p_{k}=\text{Tr}[(\Pi _{k}^{A}\otimes I_{B})\rho (\Pi _{k}^{A}\otimes I_{B})],%
\text{ }k=1,2
\end{equation}%
Recently, Rulli et al. [15] have proposed global quantum discord (QD) for
multipartite quantum states. QD for an arbitrary $N$-partite state $\rho
_{A_{1}A_{2}\text{.......}A_{N}}$ under the set of local measurement $\{\Pi
_{k}^{A_{1}}\otimes ......\otimes \Pi _{k}^{A_{N}}\}$ can be defined as%
\begin{equation}
D(\rho _{A_{1}A_{2}\text{.......}A_{N}})=\min_{\{\Pi \}}[S(\rho _{A_{1}A_{2}%
\text{.......}A_{N}}||\Phi (\rho _{A_{1}A_{2}\text{.......}%
A_{N}}))-\sum\limits_{j=1}^{N}S(\rho _{A_{j}}||\Phi _{j}(\rho _{A_{j}}))]
\end{equation}%
where $\Phi _{j}(\rho _{A_{j}})=\sum_{i}\Pi _{i}^{A_{j}}\rho _{A_{j}}\Pi
_{i}^{A_{j}}$ and $\Phi (\rho _{A_{1}A_{2}\text{.......}A_{N}})=\sum_{k}\Pi
_{k}\rho _{A_{1}A_{2}\text{.......}A_{N}}\Pi _{k}$ with $\Pi _{k}=\Pi
_{j_{1}}^{A_{1}}\otimes ......\otimes \Pi _{j_{N}}^{A_{N}}.$ One can select
the set of von-Neumann measurements as
\begin{equation}
\Pi _{1}^{A_{j}}=\left[
\begin{array}{cc}
\cos ^{2}(\frac{\theta _{j}}{2}) & e^{i\phi _{j}}\cos (\frac{\theta _{j}}{2}%
)\sin (\frac{\theta _{j}}{2}) \\
e^{-i\phi _{j}}\cos (\frac{\theta _{j}}{2})\sin (\frac{\theta _{j}}{2}) &
\sin ^{2}(\frac{\theta _{j}}{2})%
\end{array}%
\right]
\end{equation}%
\begin{equation}
\Pi _{2}^{A_{j}}=\left[
\begin{array}{cc}
\sin ^{2}(\frac{\theta _{j}}{2}) & -e^{-i\phi _{j}}\cos (\frac{\theta _{j}}{2%
})\sin (\frac{\theta _{j}}{2}) \\
e^{i\phi _{j}}\cos (\frac{\theta _{j}}{2})\sin (\frac{\theta _{j}}{2}) &
\cos ^{2}(\frac{\theta _{j}}{2})%
\end{array}%
\right]
\end{equation}%
One must find the measurement bases that minimize the QD by varying the
angles $\theta _{j}$ and $\phi _{j}$, which is achieved by adopting local
measurements in the $\sigma _{z}$ eigenbases for $GHZ$-type initial states.

The geometric quantum discord for bipartite X-state system has been proposed
by Dakic et al [58] as%
\begin{equation}
D_{G}(\rho ):=\underset{\Pi ^{A}}{\min }\left\Vert \rho -\chi \right\Vert
^{2}
\end{equation}%
where the minimum is over the set of zero-discord states $\chi $. The square
of Hilbert-Schmidt norm of Hermitian operators, $\left\Vert \rho -\chi
\right\Vert ^{2}=$Tr$[(\rho -\chi )^{2}]$. Whereas the geometric quantum
discord (GQD) for $N$-partite state $\rho _{\rho _{A_{1}A_{2}\text{.......}%
A_{N}}}$ is defined as
\begin{eqnarray}
D^{G}(\rho _{A_{1}A_{2}\text{.......}A_{N}}) &=&\sum\limits_{j=1}^{N}S(\rho
_{A_{j}})-S(\rho _{A_{1}A_{2}\text{.......}A_{N}})  \notag \\
&&-\max_{\Pi }\left[ \sum S(\Pi _{A_{j}}(\rho _{A_{i}}))-S(\Pi (\rho _{\rho
_{A_{1}A_{2}\text{.......}A_{N}}}))\right]
\end{eqnarray}%
where $\Pi =\Pi _{\rho _{A_{1}\text{.......}A_{N}}}$ is a locally projective
measurement on $A_{1}A_{2}$.......$A_{N}.$ Here in this paper, we consider
three different types of initial states as part of the following general $N$%
-qubit Werner-$GHZ$ initial state of the form

\begin{equation}
\rho =(1-\mu )\frac{I^{\otimes N}}{2^{N}}+\mu |\psi \rangle \left\langle
\psi \right\vert
\end{equation}%
where $I$ is $2\times 2$ identity operator, $\mu \in \lbrack 0,1]$ and $%
|\psi \rangle $ is the $N$-qiubit $GHZ$ state $|\psi \rangle
=(|00....0\rangle +|11.....1\rangle )/\sqrt{2}.$\newline
\textbf{(i)} Let the three partners share a three-qubit Werner-$GHZ$ initial
state as given by
\begin{equation}
\rho =(1-\mu )\frac{I^{\otimes 3}}{8}+\mu |\psi \rangle \left\langle \psi
\right\vert
\end{equation}%
where $|\psi \rangle =(|000\rangle +|111\rangle )/\sqrt{2}.$\newline
\textbf{(ii)} A six-qubit Werner-$GHZ$ initial state as given by
\begin{equation}
\rho =(1-\mu )\frac{I^{\otimes 6}}{64}+\mu |\psi \rangle \left\langle \psi
\right\vert
\end{equation}%
where $|\psi \rangle =(|000000\rangle +|111111\rangle )/\sqrt{2}.$\newline
\textbf{(iii)} Let the three observers: Alice, an inertial observer, Bob and
Charlie, the accelerated observers moving with uniform acceleration, share
the following maximally entangled $GHZ$-type initial state%
\begin{equation}
\left\vert \Psi \right\rangle _{ABC}=\frac{1}{\sqrt{2}}\left. \left(
|0_{\omega _{a}}\rangle _{A}|0_{\omega _{b}}\rangle _{B}|0_{\omega
_{c}}\rangle _{C}+|1_{\omega _{a}}\rangle _{A}|1_{\omega _{b}}\rangle
_{B}|1_{\omega _{c}}\rangle _{C}\right) \right.
\end{equation}%
where $|0_{\omega _{a(bc)}}\rangle _{A(BC)}$ and $|1_{\omega
_{a(bc)}}\rangle _{A(BC)}$ are vacuum states and the first excited states
from the perspective of an inertial observer respectively. Let the Dirac
fields, as shown in Refs. [90], from the perspective of the uniformly
accelerated observers, are described as an entangled state of two modes
monochromatic with frequency $\omega _{i},$ $\forall _{i}$%
\begin{equation}
|0_{\omega _{i}}\rangle _{M}=\cos r_{i}|0_{\omega _{i}}\rangle
_{I}|0_{\omega _{i}}\rangle _{II}+\sin r_{i}|1_{\omega _{i}}\rangle
_{I}|1_{\omega _{i}}\rangle _{II}
\end{equation}%
and the only excited state is%
\begin{equation}
|1_{\omega _{i}}\rangle _{M}=|1_{\omega _{i}}\rangle _{I}|0_{\omega
_{i}}\rangle _{II}
\end{equation}%
where $\cos r_{i}=(e^{-2\pi \omega c/a_{i}}+1)^{-1/2}$, $a_{i}$ is the
acceleration of $i^{\text{th}}$ observer. The subscripts $I$ and $II$ of the
kets represent the Rindler modes in region $I$ and $II$, respectively, in
the Rindler spacetime diagram (see Ref. [82], Fig. (1)). Considering that an
accelerated observer in Rindler region $I$ has no access to the field modes
in the causally disconnected region $II$ and by taking the trace over the
inaccessible modes, one obtains the following tripartite state in Rindler
spacetime as given by [78]%
\begin{eqnarray}
\left\vert \Psi \right\rangle _{AB_{I}C_{I}} &=&\frac{1}{\sqrt{2}}[\cos
r_{b}\cos r_{c}|0\rangle _{A}|0\rangle _{B_{I}}|0\rangle _{C_{I}}+\cos
r_{b}\sin r_{c}|0\rangle _{A}|0\rangle _{B_{I}}|1\rangle _{C_{I}}  \notag \\
&&+\sin r_{b}\cos r_{c}|0\rangle _{A}|1\rangle _{B_{I}}|0\rangle
_{C_{I}}+\sin r_{b}\sin r_{c}|0\rangle _{A}|1\rangle _{B_{I}}|1\rangle
_{C_{I}}  \notag \\
&&+|1\rangle _{A}|1\rangle _{B_{I}}|1\rangle _{C_{I}}]
\end{eqnarray}%
For the sake of simplicity, the frequency subscripts are dropped and in
density matrix formalism, the above state can be written as%
\begin{eqnarray}
\rho _{AB_{I}C_{I}} &=&\frac{1}{\sqrt{2}}[\cos r_{b}^{2}\cos
r_{c}^{2}|000\rangle \left\langle 000\right\vert +\cos r_{b}^{2}\sin
r_{c}^{2}|001\rangle \left\langle 001\right\vert   \notag \\
&&+\sin r_{b}^{2}\cos r_{c}^{2}|010\rangle \left\langle 010\right\vert +\sin
r_{b}^{2}\sin r_{c}^{2}|011\rangle \left\langle 011\right\vert   \notag \\
&&+\cos r_{b}\cos r_{c}(|000\rangle \left\langle 111\right\vert +|111\rangle
\left\langle 000\right\vert )+|111\rangle \left\langle 111\right\vert ]
\end{eqnarray}%
In order to simplify our calculations, it is assumed that Bob and Charlie
move with the same acceleration, i.e. $r_{b}=r_{c}=r.$

The interaction between the system and its environment introduces the
decoherence to the system, which is a process of the undesired correlation
between the system and the environment. The evolution of a state of a
quantum system in a noisy environment can be described by the super-operator
$\Phi $ in the Kraus operator representation as [91]

\begin{equation}
\rho _{f}=\Phi (\rho _{\text{ini}})=\sum_{k}E_{k}\rho _{i}E_{k}^{\dag }
\label{E5}
\end{equation}%
where the Kraus operators $E_{k}$ satisfy the following completeness
relation $\sum_{k}E_{k}^{\dag }E_{k}=I.$ The Kraus operators for the
evolution of $N$-partite system can be constructed from the single qubit
Kraus operators by taking their tensor product over all $n^{N}$ combinations
of $\pi \left( i\right) $ indices as $E_{k}=\underset{\pi }{\otimes }e_{\pi
\left( i\right) },$ where $n$ corresponds to the number of Kraus operators
for a single qubit channel. The single qubit Kraus operators for different
channels are given in table 1. Using equations (7-11) along with the initial
density matrices as given in equations (13, 14 and 19), the QD and GQD for
the multipartite system under different environments can be found. The
analytical relations for GQD for all the three types of initial states are
given in tables 2, 3 and 4 respectively. However, the analytical expressions
for QD are too lengthy, therefore, these are not presented in the text and
have been explained in figures.

\section{Discussions}

Analytical expressions for the quantum discord and geometric quantum discord
are calculated for different situations. Influence of different decoherence
channels such as amplitude damping, phase damping, depolarizing and phase
flip, bit flip and bit-phase flip channels is investigated for $GHZ$ type
initial states in inertial and non-inertial frames. The results consist of
three parts (i) the effect of decoherence parameter $p$ on the quantum
discord (QD) and geometric quantum discord (GQD) for three-qubit Werner-$GHZ$
states (ii) the effect of decoherence parameter $p$ on QD and GQD for
six-qubit Werner-$GHZ$ states (iii) the dynamics of quantum discord and
geometric quantum discord for three-qubit $GHZ$ state in non-inertial frames
influenced by different decoherence channels.

In figure 1, the quantum discord is plotted as a function of decoherence
parameter $p$ for $\mu =0.5$ for different noisy channels for three-qubit
and six-qubit $GHZ$ states. It is seen that the quantum discord vanishes for
$p>0.75$ in case of three-qubit $GHZ$ states and for $p>0.5$ for six qubit $%
GHZ$ states. The depolarizing channel has dominant effect on the discord as
compared to the amplitude damping channel. Whereas the behaviour of flipping
channels such as bit flip, phase flip and bit-phase flip channels is
symmetrical around 50\% decoherence as expected. It is seen that rapid
sudden death of discord occurs in case of phase flip channel. However, for
bit flip channel, no sudden death happens in case of six-qubit states. The
rise and fall of quantum discord is seen for all the flipping channels. It
is shown that states with higher number of qubits ($N=6$) are more prone to
decoherence as compared to the states with less number of qubits ($N=3$).

In figure 2, the geometric quantum discord is plotted as a function of
decoherence parameter $p$ for $\mu =0.5$ for different noisy channels, panel
(a) three-qubit $GHZ$ states and panel (b) six-qubit $GHZ$ states. It can be
seen from the figure that the depolarizing channel heavily influences the
geometric quantum discord if compared with amplitude damping channel. It is
seen that quantum discord is more robust than geometric quantum discord.
Furthermore, it is seen that the vanishing of geometric quantum discord
happens for both cases of the qubit states for bit-flip channel. In figure
3, the geometric quantum discord is plotted as a function of decoherence
parameter $p$ and acceleration $r$ (a) amplitude damping channel and (b)
bit-phase flip channel, for different values of acceleration $r$ lower
panel, respectively. It is seen that the behaviour of bit-phase flip channel
is symmetrical at $p=0.5$. It is also seen that effect of environment on
multipartite geometric quantum discord is much stronger than that of the
acceleration of non-inertial frames.

In figure 4, the quantum discord and geometric quantum discord are plotted
as a function of acceleration $r$ for different decoherence channels at $%
p=0.5$. To illustrate the environmental effects, a comparison for different
values of decoherence parameter $p$ is given in sub figure (d) for amplitude
damping channel only. It is seen that the degradation of quantum discord and
geometric quantum discord occurs due to Unruh effect. The depolarizing
channel influences the discords more heavily as compared to the other
channels in non-inertial frames. It means that the depolarizing channel has
the most destructive influence on the discords for multipartite states.

\section{Conclusions}

Quantum discord (QD) and geometric quantum discord (GQD) for multipartite
quantum states is investigated under the influence of external environments.
Different types of initial states such as, Werner-$GHZ$ type three-qubit and
six-qubit states are considered in inertial and non-inertial frames.
Dynamics of QD and GQD is investigated under amplitude damping, phase
damping, depolarizing and flipping channels. It is seen that the quantum
discord is more robust than geometric quantum discord under environmental
effects. However, Werner-$GHZ$ type states are found to be more fragile to
decoherence for higher values of $N$. Sudden death and birth of discords
occur in case of flipping channels. However, for bit flip channel, no sudden
death happens for six-qubit $GHZ$ states. Whereas depolarizing channel
heavily influences the QD and GQD as compared to the amplitude damping
channel. This implies that the depolarizing channel has the most destructive
influence on the discords for multipartite states. For the case of
accelerated observers, it is seen that the effect of environment on QD and
GQD is much stronger than that of the acceleration of the accelerated
observer. However, the degradation of QD and GQD happens due to Unruh
effect. In general, QD exhibits more robustness than GQD when multipartite
systems are exposed to environment.

\begin{figure}[tbp]
\begin{center}
\vspace{-2cm} \includegraphics[scale=0.6]{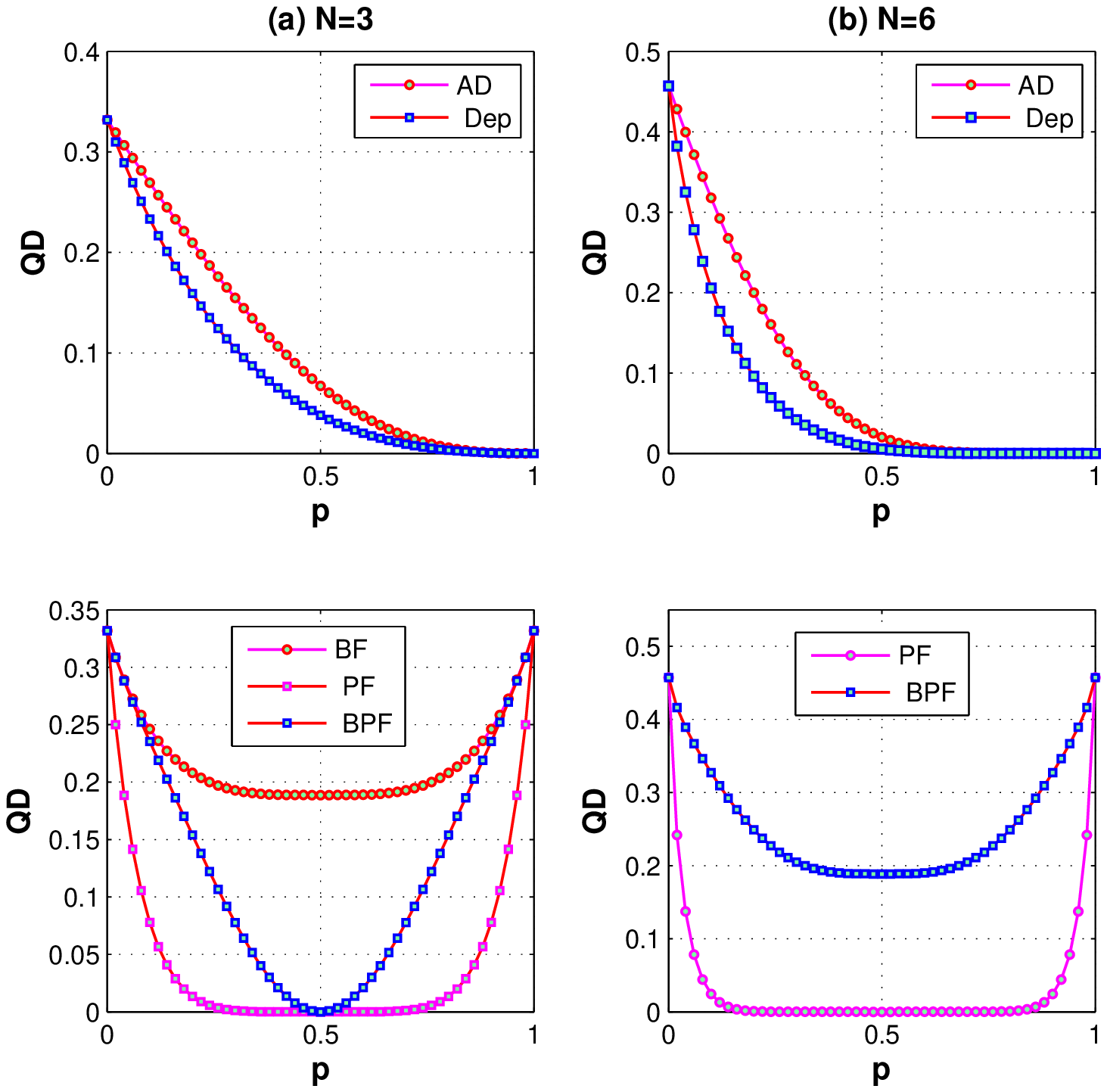} \\[0pt]
\end{center}
\caption{(Color online). Quantum discord (QD) is plotted as a function of
decoherence parameter $p$ for $\protect\mu =0.5$ for different noisy
channels, panel (a) three-qubit $GHZ$ states and panel (b) six-qubit $GHZ$
states.}
\end{figure}

\begin{figure}[tbp]
\begin{center}
\vspace{-2cm} \includegraphics[scale=0.7]{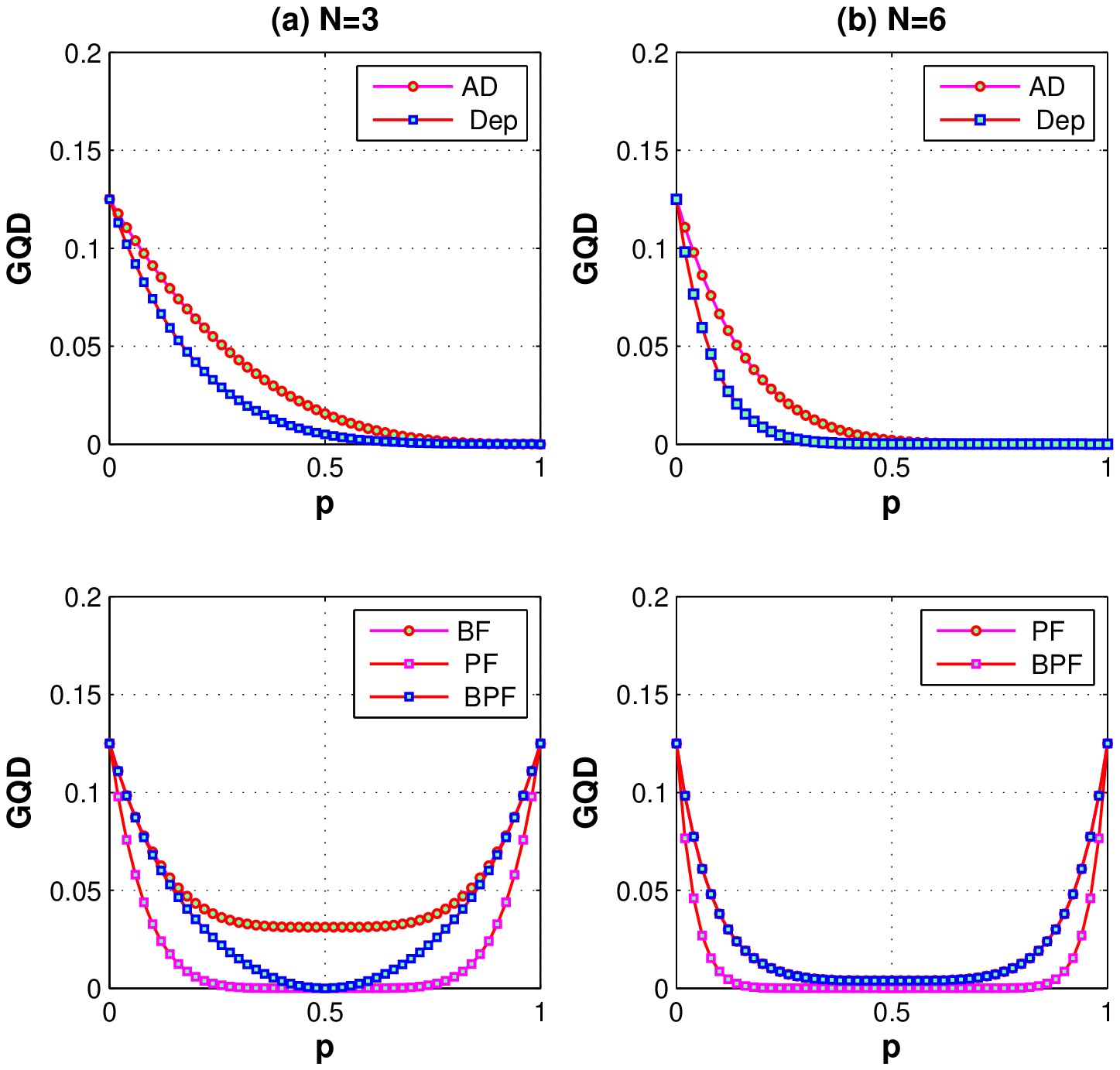} \\[0pt]
\end{center}
\caption{(Color online). Geometric quantum discord (GQD) is plotted as a
function of decoherence parameter $p$ for $\protect\mu =0.5$ for different
noisy channels, panel (a) three-qubit $GHZ$ states and panel (b) six-qubit $%
GHZ$ states.}
\end{figure}

\begin{figure}[tbp]
\begin{center}
\vspace{-2cm} \includegraphics[scale=0.6]{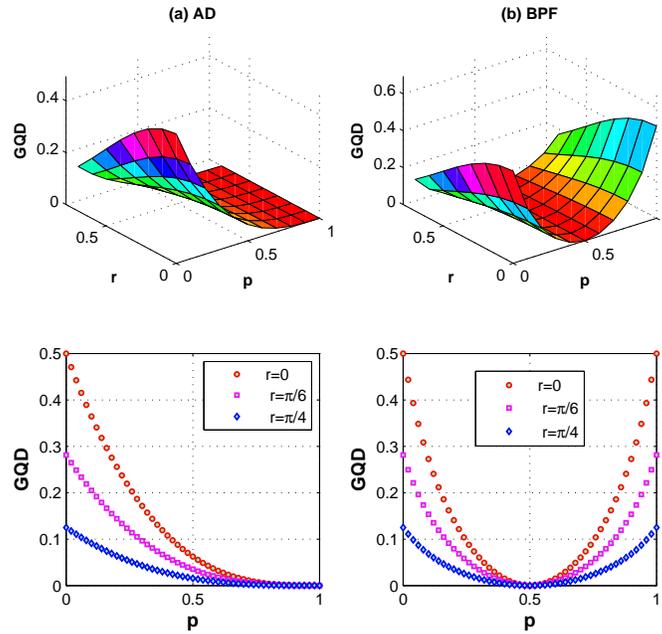} \\[0pt]
\end{center}
\caption{(Color online). Geometric quantum discord (GQD) is plotted as a
function of decoherence parameter $p$ for panel (a) amplitude damping
channel, for different values of acceleration $r$ panel (b) bit-phase flip
channel respectively.}
\end{figure}

\begin{figure}[tbp]
\begin{center}
\vspace{-2cm} \includegraphics[scale=0.6]{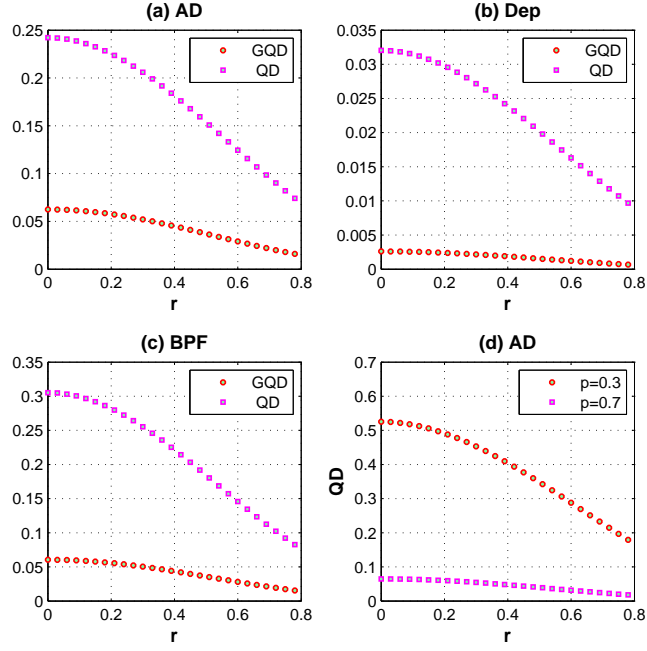} \\[0pt]
\end{center}
\caption{(Color online). Quantum discord and Geometric quantum discord are
plotted as a function of acceleration $r$ for different decoherence channels
at $p=0.5$. A comparison for different values of decoherence parameter p is
given in fig (d) for the case of amplitude damping channel.}
\end{figure}

\begin{table}[tbh]
\caption{Single qubit Kraus operators for amplitude damping, depolarizing,
phase damping, bit-phase flip, bit flip and phase flip channels where $p$
represents the decoherence parameter.}$%
\begin{tabular}{|l|l|}
\hline
&  \\
$\text{Amplitude damping channel}$ & $A_{0}=\left[
\begin{array}{cc}
1 & 0 \\
0 & \sqrt{1-p}%
\end{array}%
\right] ,$ $A_{1}=\left[
\begin{array}{cc}
0 & \sqrt{p} \\
0 & 0%
\end{array}%
\right] $ \\ \hline
$\text{Phase damping channel}$ & $E_{0}=\left[
\begin{array}{cc}
1 & 0 \\
0 & \sqrt{1-p}%
\end{array}%
\right] ,\text{ }E_{1}=\left[
\begin{array}{cc}
0 & 0 \\
0 & \sqrt{p}%
\end{array}%
\right] $ \\
$\text{Depolarizing channel}$ & $%
\begin{tabular}{l}
$A_{0}=\sqrt{1-\frac{3p}{4}I},\quad A_{1}=\sqrt{\frac{p}{4}}\sigma _{x}$ \\
$A_{2}=\sqrt{\frac{p}{4}}\sigma _{y},\quad \quad $\ $\ A_{3}=\sqrt{\frac{p}{4%
}}\sigma _{z}$%
\end{tabular}%
$ \\
&  \\ \hline
$\text{Bit-phase flip channel}$ & $A_{0}=\sqrt{1-p}I,\quad A_{1}=\sqrt{p}%
\sigma _{y}$ \\
&  \\ \hline
$\text{Bit flip channel}$ & $A_{0}=\sqrt{1-p}I,\quad A_{1}=\sqrt{p}\sigma
_{x}$ \\
&  \\ \hline
$\text{Phase flip channel}$ & $A_{0}=\sqrt{1-p}I,\quad A_{1}=\sqrt{p}\sigma
_{z}$ \\ \hline
\end{tabular}%
$%
\label{di-fit}
\end{table}
\begin{table}[tbh]
\caption{Analytical expressions of GQD for Werner-$GHZ$ three-qubit state
under different environments.}%
\begin{tabular}{|l|l|}
\hline
Channel Description & GQD (3-qubit $GHZ$\ state) \\ \hline
$\text{Amplitude damping}$ & $\frac{1}{2}\left( 1-p\right) ^{3}\mu ^{2}$ \\
&  \\ \hline
Depolarizing & $%
\begin{tabular}{l}
$\frac{1}{512}\left( 4-3\,p\right) ^{4}\left( 1-p\right) ^{2}\mu ^{2}$ \\
$+\frac{1}{512}p^{2}\left( 4+p\left( 7-3\,p\right) \right) ^{2}\mu ^{2}$%
\end{tabular}%
$ \\
&  \\ \hline
Phase damping & $\frac{1}{2}\left( 1-p\right) ^{3}\mu ^{2}$ \\
&  \\ \hline
Bit flip & $%
\begin{tabular}{l}
$\frac{3}{2}p^{2}\left( 1-3p+2\,p^{2}\right) ^{2}\mu ^{2}$ \\
$+\frac{1}{2}\left( 1-p\left( 3-3p+2p^{2}\right) \right) ^{2}\mu ^{2}$%
\end{tabular}%
$ \\
&  \\ \hline
$\text{Phase flip}$ & $\frac{1}{2}\left( 1-2p\right) ^{6}\mu ^{2}$ \\
&  \\ \hline
Bit-phase flip & $%
\begin{tabular}{l}
$\frac{3}{2}p^{2}\left( 1-3p+2\,p^{2}\right) ^{2}\mu ^{2}$ \\
$+\frac{1}{2}\left( 1-p\left( 3-3p+2p^{2}\right) \right) ^{2}\mu ^{2}$%
\end{tabular}%
$ \\ \hline
\end{tabular}%
\end{table}
\begin{table}[tbh]
\caption{Analytical expressions of GQD for Werner-$GHZ$ six-qubit state
under different environments.}%
\begin{tabular}{|l|l|}
\hline
Channel Description & GQD (6-qubit $GHZ$\ state) \\ \hline
$\text{Amplitude damping}$ & $\frac{1}{2}\left( 1-p\right) ^{6}\mu ^{2}$ \\
&  \\ \hline
Depolarizing & $\frac{1}{2}\left( 1-p\right) ^{12}\mu ^{2}$ \\
&  \\ \hline
Phase damping & $\frac{1}{2}\left( 1-p\right) ^{6}\mu ^{2}$ \\
&  \\ \hline
Bit flip & $%
\begin{tabular}{l}
$20\left( 1-p\right) ^{6}p^{6}\mu ^{2}+\frac{15}{2}\left( 1-p\right)
^{4}p^{4}\times $ \\
$\left( 1-2p+2p^{2}\right) ^{2}\mu ^{2}+\frac{1}{2}\left( \left(
1-p\right) ^{6}+p^{6}\right) ^{2}\mu ^{2}$ \\
$+3\left( 1-p\right) ^{2}p^{2}\left( 1-4p+6p^{2}-4p^{3}+2p^{4}\right)
^{2}\mu ^{2}$%
\end{tabular}%
$ \\
&  \\ \hline
$\text{Phase flip}$ & $\frac{1}{2}\left( 1-2p\right) ^{12}\mu ^{2}$ \\
&  \\ \hline
Bit-phase flip & $%
\begin{tabular}{l}
$20\left( 1-p\right) ^{6}p^{6}\mu ^{2}+\frac{15}{2}\left( 1-p\right)
^{4}p^{4}\times $ \\
$\left( 1-2p+2p^{2}\right) ^{2}\mu ^{2}+\frac{1}{2}\left( \left(
1-p\right) ^{6}+p^{6}\right) ^{2}\mu ^{2}$ \\
$+3\left( 1-p\right) ^{2}p^{2}\left( 1-4p+6p^{2}-4p^{3}+2p^{4}\right)
^{2}\mu ^{2}$%
\end{tabular}%
$ \\ \hline
\end{tabular}%
\end{table}
\begin{table}[tbh]
\caption{Analytical expressions of GQD for $GHZ$ state in non-inertial
frames (Eq. 19) under different environments.}%
\begin{tabular}{|l|l|}
\hline
Channel Description & GQD (3-qubit state Eq. (19)) \\ \hline
$\text{Amplitude damping}$ & $\frac{1}{2}\left( 1-p\right) ^{3}\cos ^{4}(r)$
\\
&  \\ \hline
Depolarizing & $%
\begin{tabular}{l}
$\frac{1}{512}\left( 4-3p\right) ^{4}\left( 1-p\right) ^{2}\cos ^{4}\left(
r\right) $ \\
$+\frac{1}{512}p^{2}(4-7p+3p^{2})\cos ^{4}\left( r\right) $%
\end{tabular}%
$ \\
&  \\ \hline
Phase damping & $\frac{1}{2}\left( 1-p\right) ^{3}\cos ^{4}(r)$ \\
&  \\ \hline
$\text{Phase flip}$ & $\frac{1}{2}\left( 1-2p\right) ^{6}\cos ^{4}(r)$ \\
&  \\ \hline
Bit-phase flip & $%
\begin{tabular}{l}
$\frac{3}{2}p^{2}\left( 1-3p+2p^{2}\right) ^{2}\cos ^{4}(r)$ \\
$+\frac{1}{2}\left( 1-3p+3p^{2}-2p^{3}\right) ^{2}\cos ^{4}(r)$%
\end{tabular}%
$ \\ \hline
\end{tabular}%
\end{table}


\begin{thebibliography}{99}
\bibitem{1} H. Ollivier,and W. H. Zurek, Phys. Rev. Lett. \textbf{88},
017901 (2001).

\bibitem{2} A. Datta, A. Shaji, and C. M. Caves, Phys. Rev. Lett. \textbf{100%
}, 050502 (2008).

\bibitem{3} S. Luo, Phys. Rev. A \textbf{77}, 042303 (2008).

\bibitem{4} M. S. Sarandy, Phys. Rev. A \textbf{80}, 022108 (2009).

\bibitem{5} A. Shabani and D. A. Lidar, Phys. Rev. Lett. \textbf{102},
100402 (2009).

\bibitem{6} M. Ali, A. R. P. Rau and G. Alber, Phys. Rev. A \textbf{81},
042105\textbf{\ }(2010).

\bibitem{7}  P. Giorda and M. G. A. Paris, Phys. Rev. Lett. \textbf{105},
020503 (2010).

\bibitem{8} J-s. Jin, \textit{et al}. J. Opt. Soc. Am. B \textbf{27}, 1799
(2010).

\bibitem{9} A. Brodutch and D. R. Terno, Phys. Rev. A \textbf{83}, 010301(R)
(2011).

\bibitem{10} Q. Chen, C. Zhang, S. Yu, X.X. Yi and C.H. Oh, Phys. Rev. A
\textbf{84}, 042313\textbf{\ }(2011).

\bibitem{11} I. Chakrabarty, P. Agrawal and A. K. Pati, Eur. Phys. J. D
\textbf{65}, 605 (2011).

\bibitem{12} A. Streltsov \textit{et al}. Phys. Rev. Lett. \textbf{106},
160401 (2011).

\bibitem{13} Jianwei Xu, J. Phys. A: Math. Theor. \textbf{44}, 445310 (2011).

\bibitem{14} B. Dakic \textit{et al.} e-print arxiv:1203.1629.

\bibitem{15} C. C. Rulli and M. S. Sarandy Phys. Rev. A \textbf{84}, 042109
(2011).

\bibitem{16} M. Okrasaa and Z. Walczak, Eur. Phys. Lett. \textbf{96,} 60003
(2011).

\bibitem{17} K. Modi and V. Vedral, AIP Conf. Proc. \textbf{1384}, 69-75
(2011).

\bibitem{18} Jianwei Xu, e-print arxiv:1204.5868.

\bibitem{19} T. Werlang, S. Souza, F. F. Fanchini and C. J. V. Boas, Phys.
Rev. A \textbf{80} 024103 (2009).

\bibitem{20} J. Maziero, L. C. C%
\'{}%
eleri, R. M. Serra, and V. Vedral, Phys. Rev. A \textbf{80}, 044102 (2009).

\bibitem{21} F. F. Fanchini, T.Werlang, C. A. Brasil, L. G. E. Arruda, and
A. O. Caldeira, Phys. Rev. A \textbf{81}, 052107 (2010).

\bibitem{22} J. Maziero, T. Werlang, F. F. Fanchini, L. C. C%
\'{}%
eleri, and R. M. Serra, Phys. Rev. A \textbf{81}, 022116 (2010).

\bibitem{23} F. F. Fanchini \textit{et al}. Phys. Rev. A \textbf{81,} 052107
(2010).

\bibitem{24} B. Wang, Z. Y Xu, Z. Q. Chen and M. Feng, Phys. Rev. A \textbf{%
81,} 014101 (2010).

\bibitem{25} A. Ferdi, Opt. Commun. \textbf{283,} 5264 (2010).

\bibitem{26} H. Zhi, J. Zou, B. Shao, S.-Y. Kong, J. Phys. B: At. Mol. Opt.
Phys. \textbf{43,} 115503 (2010).

\bibitem{27} Y.Q. Zhang and J.B. Xu, Eur. Phys. J. D \textbf{64}, 549 (2011).

\bibitem{28} A. Isar, Open Sys. \& Inf. Dyn. \textbf{18}, 175 (2011).

\bibitem{29} J. Batle \textit{et al.} J. Phys. A: Math. Theor. \textbf{44,}
505304 (2011).

\bibitem{30} B.-F. Ding, \textit{et al}. Chin. Phys. Lett. \textbf{28,}
104216 (2011).

\bibitem{31} X. Zhengjun \textit{et al. }J. Phys. B: At. Mol. Opt. Phys.
\textbf{44,} 215501 (2011).

\bibitem{32} Z. Y. Xu, \textit{et al}., J. Phys. A: Math. Theor. \textbf{44,}
395304 (2011).

\bibitem{33} S. M. Xiao, \textit{et al}., Opt. Commun. \textbf{284,} 555
(2011).

\bibitem{34} J.-Q. Li, J.-Q. Liang, Phys. Lett. A \textbf{375,} 1496 (2011).

\bibitem{35} G. Karpat, Z. Gedik, Phys. Lett. A \textbf{375,} 4166 (2011).

\bibitem{36} B. Bellomo \textit{et al}. Int. J. Quant. Inf. \textbf{9}, 1665
(2011).

\bibitem{37} K Berrada \textit{et al}. J. Phys. B: At. Mol. Opt. Phys. 44
145503 (2011).

\bibitem{38} J. L. Guo, Y. J. Mi, H. S. Song, Eur. Phys. J. D \textbf{66,
24\ }(2012).

\bibitem{39} Q-X. Mu, Y-Q. Zhang, J. Song, J. Mod. Opt. \textbf{59}, 387
(2012).

\bibitem{40} Q. Yi, J.-B. XU, Chin. Phys. Lett. \textbf{29,} 040302 (2012).

\bibitem{41} Y.-J. Mi, Int. J. Theor. Phys. \textbf{51}, 544 (2012).

\bibitem{42} A. Kofman, Quant. Info. Proc., \textbf{11,} 269 (2012).

\bibitem{43} F. Benatti, R. Floreanini, U. Marzolino, Ann. Phys. \textbf{327,%
} 1304 (2012).

\bibitem{44} Zhihua Guo et al., J. Phys. A: Math. Theor. \textbf{45}, 145301
(2012).

\bibitem{45} M. Mahdian, R. Yousefjani, S. Salimi, e-print arXiv:1204.1217.

\bibitem{46} M. Ali, J. Phys. A: Math. Theor. \textbf{43,} 495303 (2010).

\bibitem{47} S. Vinjanampathy, A. R. P. Rau, J. Phys. A: Math. Theor.
\textbf{45,} 095303 (2012).

\bibitem{48} R. Dillenschneider, Phys. Rev. B \textbf{78,} 224413 (2008).

\bibitem{49} T. Zehua, J. Jiliang, Phys. Lett. B \textbf{707,} 264 (2012).

\bibitem{50} A. Datta, Phys. Rev. A \textbf{80,} 052304 (2009).

\bibitem{51} Y. Yao, \textit{et al}. Phys. Lett. A \textbf{376,} 358 (2012).

\bibitem{52} X. Zhengjun X.-M. Lu, X. Wang, Y. Li, J. Phys. A: Math. Theor.
\textbf{44,} 375301 (2011).

\bibitem{53} S. Rana, P. Parashar, Phys. Rev. A \textbf{85}, 024102 (2012).

\bibitem{54} B. P. Lanyon \textit{et al.} Phys. Rev. Lett. 101, 200501 (2008)

\bibitem{55} G. Passante \textit{et al}. Phys. Rev. A. \textbf{84}, 044302
(2011).

\bibitem{56} A. Chiuri, \textit{et al.} Phys. Rev. A \textbf{84}, 020304(R)
(2011).

\bibitem{57} L. C. C\'{e}leri, \textit{et al.} Int. J. Quant. Inf. \textbf{9}%
, 1837 (2011).

\bibitem{58} B. Dakic \textit{et al}. Phys. Rev. Lett. \textbf{105},190502
(2010).

\bibitem{59} S.-L. Luo, S.-S. Fu, Phys. Rev. A \textbf{82,} 034302 (2010).

\bibitem{60} J. Xu, Phys. Lett. A \textbf{376}, 320 (2012).

\bibitem{61} J. Xu, e-print arxiv:1205.0330.

\bibitem{62} M-L. Hu, H. Fan, Ann. Phys. \textbf{327,} 851 (2012).

\bibitem{63} X-M. Lu, \textit{et al.} Quant. Inf. Comp. \textbf{10,} 0994
(2010).

\bibitem{64} W. H. Zurek, Rev. Mod. Phys. \textbf{75}, 715 (2003).

\bibitem{65} M. A. Schlosshauer, Decoherence and the Quantum-To- Classical
Transition (Springer, 2007).

\bibitem{66} M. Brune, \textit{et al.} Phys. Rev. Lett. \textbf{77}, 4887
(1996)

\bibitem{67} S. Scheel, D.-G. Welsch, Phys. Rev. A \textbf{64,} 063811
(2001).

\bibitem{68} Q. Pan, J. Jing, Phys. Rev. A \textit{77,} 024302 (2008).

\bibitem{69} M. Ramzan, M. K. Khan, Quant. Inf. Proc. \textbf{9,} 667 (2010).

\bibitem{70} Y-S. Kim, \textit{et al.} Nature Phys. \textbf{8,} 117 (2012).

\bibitem{71} P. M. Alsing, I. Fuentes-Schuller, R. B. Mann, T.E. Tessier,
Phys. Rev. A \textbf{74,} 032326\ (2006).

\bibitem{72} D. E. Bruschi, \textit{et al.} Phys. Rev. A \textbf{82,} 042332
(2010).

\bibitem{73} J. Wang, J. Deng, J. Jing, Phys. Rev. A \textbf{81,} 052120
(2010).

\bibitem{74} E. Martn-Martnez, et al., Phys. Rev. D \textbf{82,} 064006
(2010).

\bibitem{75} M. R. Hwang, \textit{et al.} Phys. Rev. A \textbf{83,} 012111
(2010).

\bibitem{76} J. Wang, J. Jing, Phys. Rev. A \textbf{83,} 022314 (2011).

\bibitem{77} M. Montero, \textit{et al.} Phys. Rev. A \textbf{84,} 042320
(2011).

\bibitem{78} M.-R. Hwang, D. Park, E. Jung, Phys. Rev. A \textbf{83,} 012111
(2010).

\bibitem{79} M-D. Hossein, \textit{et al}. Ann. Phys. \textbf{326,} 1320
(2011).

\bibitem{80} T. C. Ralph and T. G. Downes, Cont. Phys. \textbf{53}, 1 (2012).

\bibitem{81} L.C. Celeri,\ \textit{et al.} Phys. Rev. A \textbf{81,} 062130
(2010).

\bibitem{82} J. Wang, J. Jing, Phys. Rev. A \textbf{82,} 032324 (2010).

\bibitem{83} J. Wang, J. Jing, Ann. Phys. \textbf{327,} 283 (2012).

\bibitem{84} Min-Zhe Piao, Xin Ji, J. Mod. Opt. \textbf{59}, 21 (2012).

\bibitem{85} Y. Wang, Xin Ji, J. Mod. Opt. \textbf{59}, 571 (2012).

\bibitem{86} M. Ramzan, M. K. Khan, Quant. Inf. Process. \textbf{11}, 443
(2012).

\bibitem{87} M. Ramzan, Chin. Phys. Lett. \textbf{29,} 020302 (2012).

\bibitem{88} M. Ramzan, Quant. Inf. Proc. DOI:10.1007/s11128-011-0354-7
(2012).

\bibitem{89} M. Ramzan, e print arXiv:1204.1900.

\bibitem{90} M. Aspachs, \textit{et al}. Phys. Rev. Lett \textbf{105} 151301
(2010).

\bibitem{92} M. A. Nielson and I. L. Chuang, Quantum Computation and Quantum
Information (Cambridge University Press, Cambridge, 2000).\pagebreak
\end{thebibliography}
\end{document}